\documentclass[aps,prb,twocolumn,floatfix,showpacs,superscriptaddress]{revtex4}

\usepackage{graphicx}
\usepackage{bbm}
\usepackage{amsmath,amssymb}

\newcommand{\be}{\begin{equation}}
\newcommand{\beq}{\begin{equation}}
\newcommand{\ee}{\end{equation}}
\newcommand{\bea}{\begin{eqnarray}}
\newcommand{\eea}{\end{eqnarray}}
\newcommand{\ba}{\begin{array}}
\newcommand{\ea}{\end{array}}

\renewcommand{\vr} {{\bf r}}

\newcommand{\vs} {{\bf s}}

\newcommand{\vj} {{\bf j}}

\newcommand{\nn} {\nonumber}

\begin{document}
\title{Laplacian-level density functionals for the exchange-correlation energy of low-dimensional nanostructures}
\author{S. Pittalis}
\email[Electronic address:\;]{pittaliss@missouri.edu}
\affiliation{Department of Physics and Astronomy, University of Missouri, Columbia, Missouri 65211, USA}
\author{E. R{\"a}s{\"a}nen}
\email[Electronic address:\;]{erasanen@jyu.fi}
\affiliation{Nanoscience Center, Department of Physics, University of
  Jyv\"askyl\"a, FI-40014 Jyv\"askyl\"a, Finland}

\date{\today}

\begin{abstract}
In modeling low-dimensional electronic nanostructures, the evaluation
of the electron-electron interaction is a challenging task. Here we
present an accurate and practical density-functional approach to the
two-dimensional many-electron problem. In particular, we show that
spin-density functionals in the class of meta-generalized-gradient
approximations can be greatly simplified by reducing the explicit
dependence on the Kohn-Sham orbitals to the dependence on the electron
spin density and its spatial derivatives. Tests on various quantum-dot
systems show that the overall accuracy is well preserved, if not even
improved, by the modifications.
\end{abstract}

 
\maketitle

\section{Introduction}

With the present technology the
electron gas can be confined in various ways to create
nanoscale devices of lower dimension. The
field of two-dimensional (2D) physics has grown rapidly 
alongside the development of electronic devices such as
quantum Hall bars and point contacts, and semiconductor quantum 
dots. When modeling these systems, the finite extent
in the growth direction (say $z$) can often be neglected,
so that the system is well described by a 2D Hamiltonian
in the effective-mass approximation.~\cite{qd} In this respect,
the building block is the 2D electron gas (2DEG), whose
properties are well known in the literature.~\cite{vignalebook}

Density-functional theory (DFT) and its extensions 
have become the method of choice to describe
the electronic properties of three-dimensional (3D) systems
such as atoms, molecules and solids.~\cite{dft1,dft3}
Despite the fact that the many 3D density functionals developed 
for the exchange-correlation (xc) energy and potential
fail in the quasi-2D limit,~\cite{kim,pollack,cost1,cost2} the
derivation of {\em explicitly 2D} xc functionals has started 
only very recently.~\cite{x1,ring,gamma,c1,c2,bj,lccs,dga,of} 
In finite 2D systems, most of these 
functionals overperform the 2D local spin-density
approximation (LSDA), which is a combination of the analytic
exchange energy of the 2DEG~\cite{rajagopal} 
and the corresponding correlation energy having 
a parametrized form.~\cite{tanatar,attaccalite}
The encouraging results obtained with the new
functionals indicate that
DFT in 2D has entered in a more mature phase.  

Among the newly proposed functionals, our focus is on those 
approximations that have as ingredients the electron density and 
its spatial derivatives, the kinetic energy density
and the paramagnetic current.~\cite{x1,ring,dga,c1,c2,gamma,bj}  
In other words, the expressions of these functionals are 
current-dependent meta-generalized-gradient approximations (meta-GGAs),
and therefore they explicitly depend on the Kohn-Sham (KS) orbitals.
As it is valid for 3D systems, also for 2D systems the 
meta-GGAs are very accurate. However, the price for their accuracy 
is the more involved numerical implementation (if applied
self-consistently) as well as the (case-dependent) 
numerical burden in the applications.
In order to simplify both the mentioned tasks, we explore here
to which extent and how the explicit dependence on the KS 
orbitals may be reduced to the dependence on the electron density
and its spatial derivatives, yet possibly maintaining a 
satisfactory level of accuracy. In other words, we examine the
path {\em from implicit to explicit} density functionals in 
the class of meta-GGAs.
A similar study has been already carried out for 3D systems,
introducing some Laplacian-level meta-GGA functionals.~\cite{LapLev}
With the present work, we explore this possibility for 
low-dimensional systems. We find that the performance after
the simplifications is well preserved, and in some cases 
even improved.

\section{Review and modifications of the functionals}

In the following, we review the main ingredients of recently derived
functionals and suggest how their expressions may be significantly
simplified in a consistent fashion. Ideally,
the ultimate goal is to obtain the best performance
with the least (numerical) effort. In particular, we 
consider (i) an exchange-energy functional 
obtained through the modeling of the exchange-hole (x-hole) 
functions,~\cite{x1} (ii) an exchange-energy functional 
obtained from the one-body-density-matrix~\cite{gamma} (1BSDM),
and (iii) correlation-energy functionals
obtained through the modeling of the correlation hole.~\cite{c1,c2}
Details of the derivation of the functionals can be found the in the 
mentioned literature.

\subsection{Exchange energy from the exchange-hole functions}\label{x-hole}

The exchange energy $E_x$ can be expressed through the x-hole 
functions as~\cite{dft1,dft3} $h^{\sigma}_x(\vr_1,\vr_2)$
\begin{equation} \label{ExH}
E_x =  \frac{1}{2}  \sum_{\sigma} \int d^2 r \rho_{\sigma}(\vr)  \int_{0}^{\infty}  ds \,
\int_0^{2\pi} d\phi_s~h^{\sigma}_x(\vr,\vr+\vs),
\end{equation}
with
\begin{equation}
h^{\sigma}_{x}(\vr_1,\vr_2) = -
\frac{|\sum_{k=1}^{N_\sigma}\psi^*_{k,\sigma}(\vr_1)\psi_{k,\sigma}(\vr_2)|^2}
{\rho_{\sigma}(\vr_1)}\;,
\label{xhole}
\end{equation}
where $\psi_{k\sigma}(\vr)$ are the KS (spin) orbitals.
From Eq.~(\ref{ExH}), it is apparent that an approximation 
of the x-hole also provides an approximation for the exchange energies.
Equation~(\ref{ExH}) also suggests that the details of the 
angular dependence of the x-hole are energetically negligible:
all we need is the {\em cylindrical} average of the x-hole, {\em i.e.}
$\bar{h}^{\sigma}_x(\vr;s)$. 

As the basis of our model,~\cite{x1}  we have chosen
\begin{eqnarray}\label{h}
\bar{h}^{\sigma}_x(\vr;s) &=&  \frac{1}{2\pi}   \int_0^{2\pi} d\phi_s~ h^{\sigma}_x(\vr;\vr+\vs) \nn \\
&\approx&  -\frac{a}{\pi} \exp \left[-a(\vr) \left( b(\vr) + s^2 \right) \right] \nn \\
& \times & I_0\left(2a(\vr) \sqrt{b(\vr)}s\right)\;,
\end{eqnarray}
where $I_0(x)$ is the zeroth-order modified Bessel function of the first kind. 
This model provides the correct sign of the x-hole and the correct 
normalization $\int d^2 s~h^{\sigma}_x(\vr,\vs) = -1$.
The non-negative functions $a(\vr)$ and $b(\vr)$ are introduced 
to reproduce the short-range behavior of the x-hole. 
It is important to note that the curvature
of the x-hole in 2D is given by~\cite{x1}
\begin{equation}\label{C}
C^{\sigma}_x(\vr) = \frac{1}{4}\left[ \nabla^2 \rho_{\sigma}(\vr) -2\tau_\sigma(\vr)
+ \frac{1}{2}\frac{\left( \nabla \rho_\sigma(\vr) \right)^2}{\rho_\sigma(\vr)}
+ 2 \frac{\vj^2_{p,\sigma}(\vr)}{\rho_\sigma(\vr)} \right]\;,
\end{equation}
where $\tau_\sigma$ is (twice) the spin-dependent kinetic-energy 
density, and $\vj_{p,\sigma}$ is the spin-dependent paramagnetic 
current density.

In  Ref.~\onlinecite{x1}, we have applied the above 
scheme in two ways. 
In the first instance, we have employed it in its full spirit
by numerically determining $a$ and $b$ at each point in space. 
Then, in a fully self-consistent application, 
the scheme should be implemented within the 
optimized-effective-potential (OEP) 
method.~\cite{SharpHorton:53,TalmanShadwick:76,KummelKronik, Engel:03,GraboKreibichKurthGross:00}
This is necessary because $\tau_\sigma$  and $\vj_{p,\sigma}$ 
(in non-current-spin-density functional
calculations~\cite{VR1,VR2}) 
make the overall scheme explicitly dependent on the 
KS orbitals.
As a consequence, the corresponding numerical task 
would be highly nontrivial. 

In the second instance, we have analyzed the 2DEG 
limit, for which
\begin{equation}\label{Ch}
C^{\sigma}_{x} (\vr)  \rightarrow  C^{\sigma}_{h,x} (\vr) = - \pi  \rho^2_{\sigma}(\vr)\;.
\end{equation}
In this way, a {\em local} density functional has been recovered,
which was seen to improve over the exchange energies obtained 
within the standard LSDA when applied to few-electron quantum dots.

Next we proceed from the review of the functional to its 
modification. As discussed above, the task is to remove the
explicit reference to the KS orbitals,  yet retaining some degree 
of flexibility in dealing with inhomogeneous systems.  
This may be achieved by introducing the following modification:~\cite{Berkane,Zyl2,dga}
\be
  \label{te}
  \tau_{\sigma}(\vr) = \sum_{k=1}^{N_\sigma} \psi_{k\sigma}(\vr) \rightarrow  \tilde{\tau}_{\sigma}(\vr) = 2\pi \rho^2_{\sigma}(\vr)+ \frac{1}{3} \nabla^2 \rho_{\sigma}(\vr) + 
  \frac{\vj^2_{p,\sigma}(\vr)}{\rho_\sigma(\vr)}\,\;.
\ee
Correspondingly, we obtain
\begin{equation}\label{C2}
C^{\sigma}_x(\vr) \rightarrow  \tilde{C}^{\sigma}_{x} (\vr) = - \pi  \rho^2_{\sigma}(\vr) + \frac{1}{12} \nabla^2  \rho_{\sigma}(\vr) 
+ \frac{1}{8}\frac{\left( \nabla \rho_\sigma(\vr) \right)^2}{\rho_\sigma(\vr)}\;.
\end{equation}
It is worth mentioning that due to the last (current-dependent) term 
on the right hand side of Eq.~(\ref{te}), 
the modified x-hole curvature manifestly preserves its gauge invariance.
As described in Ref.~\onlinecite{x1}, the functions $a$ and $b$ are 
determined from
\be
a = \pi\rho_{\sigma}\exp{(y)},
\label{a}
\ee
and
\be
b = \frac{y}{\pi\rho_{\sigma}}\exp{(-y)}\;\,
\label{b}
\ee
where $y=ab$ satisfies
\begin{equation}
\left(y-1\right)\exp(y) = \frac{\tilde{C}^{\sigma}_{x}}{\pi\rho^2_{\sigma}} = -1 + \frac{1}{12 \pi} \frac{\nabla^2 \rho_\sigma}{\rho^2_\sigma} 
+ \frac{1}{8\pi} \frac{\left( \nabla \rho\right)^2}{\rho^3_\sigma} \;.
\label{relation}
\end{equation}
If a solution does not exist, we set~\cite{x1} $y \equiv 0$. This 
corresponds to the 2DEG limit mentioned just above.
From Eq.~(\ref{relation}), it is apparent that the second 
and third terms are relevant when the curvature and
the gradient of the spin-density are non-negligible. 

Going back to the first modification 
$\tau_\sigma\rightarrow{\tilde \tau}_\sigma$,
we refer to the new (angular-averaged) 
x-hole function as $\tilde{\bar{h}}^{\sigma}_x(\vr,s)$. 
The corresponding x-hole potentials denoted as 
$\tilde{U}^\sigma_{x,{\rm model}}(\vr)$, 
read as follows~\cite{x1} 
\begin{equation}
\tilde{U}^\sigma_{x,{\rm model}}(\vr) = 2 \pi \int_{0}^{\infty}  ds \,\tilde{\bar{h}}^{\sigma}_x(\vr,s)\;
\label{xholepot1}
\end{equation}
from which, the exchange energy is obtained as
\begin{equation}
\tilde{E}_x^{\rm model} = \frac{1}{2} \sum_{\sigma} \int d^2 r \rho_\sigma(\vr) \tilde{U}^\sigma_x(\vr)\;.
\label{xenergytilde}
\end{equation}
The last two quantities calculated {\em without} the modification
of $\tau_\sigma$ are denoted below simply without the tilde 
symbols, i.e., ${U}^\sigma_{x,{\rm model}}$ and
${E}^\sigma_{x,{\rm model}}$.

Finally, we point out that the above modifications
offer a straightforward way to calculate
the corresponding KS exchange potential
as a a functional derivative, 
$v^{\sigma}_x=\delta\tilde{E}_x/(\delta\rho_\sigma)$.
The properties and performance of these 
potentials will be assessed elsewhere.

\subsection{Exchange energy from the 
one-body-spin-density matrix}\label{1BSDMsec}

Another way to express the exchange energy is to make use of the 1BSDM
$\gamma_{\sigma}\left(\vr_1,\vr_2\right)$,~\cite{dft1,dft3} so that
\begin{eqnarray}
E_x 
&=&  - \frac{1}{2 } \sum_{\sigma=\uparrow,\downarrow} 
\int d^2r \int_{0}^{\infty} \,ds  \nn\\ 
&\times& \int_{0}^{2\pi} d\phi_s
\Big|\gamma_{\sigma}\left(\vr+\frac{\vs}{2},\vr - \frac{\vs}{2}\right)\Big|^2 
\label{EX_3}
\end{eqnarray}
with
\begin{equation}
\gamma_{\sigma}(\vr_1,\vr_2) =
\sum_{k=1}^{N_\sigma}\psi_{k,\sigma}(\vr_1)\psi^*_{k,\sigma}(\vr_2)\;,
\label{gamma}
\end{equation}
where $\psi_{k\sigma}(\vr)$ are the KS (spin) orbitals.
Clearly, an approximation for the 1BSDM implies an approximation
for the exchange energy.  Also, it is apparent that
the angular dependence of the 1BSDM is energetically of minor
importance. Therefore, as a basis of our approximation 
we have considered the following expression~\cite{gamma}
\begin{eqnarray}
 \frac{1}{2\pi}  \int_{0}^{\infty} d\phi_s
\Big|\gamma_{\sigma}\left(\vr+\frac{\vs}{2},\vr - \frac{\vs}{2}\right)\Big|^2 &\approx& 
\rho_{\sigma}^2(\vr) e^{-\frac{s^2}{\beta_{\sigma}(\vr)}} \nn \\
\times \left\{ 1 +  \left[ \frac{s}{\beta_{\sigma}(\vr)} \right]^2  A_\sigma(N_\sigma)
\right\} \;,
\label{AG_6}
\end{eqnarray}
where $\beta(\vr)$ is chosen to reproduce the 
exact-short behavior of the 1BSDM
\begin{equation}
\beta_{\sigma}^{-1} (\vr)= \frac{1}{2}\frac{\tau_{\sigma}(\vr)}{\rho_\sigma(\vr)} 
- \frac{1}{8} \frac{\nabla^2\rho_{\sigma}(\vr)}{\rho_\sigma(\vr)}
-\frac{1}{2}\left(\frac{\vj_{p,\sigma}(\vr)}{\rho_{\sigma}(\vr)}\right)^2
\label{temp1}
\end{equation}
and $A_\sigma(N_\sigma)$ is obtained through the normalization
of the particle number for each spin channel.
In Ref.~\onlinecite{gamma} we have employed this scheme 
leading to accurate results for various quantum-dot systems.
In addition, we have observed that this 
approach and the one of Sec.~\ref{x-hole} 
coincide in the 2DEG limit.

Here we suggest to simplify the present functional by making 
use of Eq.~(\ref{te}) in Eq.~(\ref{temp1}). This yields
\begin{equation}
\tilde{\beta}_{\sigma}^{-1} (\vr)= \pi \rho_{\sigma}(\vr)
+ \frac{1}{24} \frac{\nabla^2\rho_{\sigma}(\vr)}{\rho_\sigma(\vr)}\;.
\label{temp2}
\end{equation}
We emphasize that no gradients of the 
spin-density appear in this expression. The resulting expression is
manifestly  gauge invariant. The relevance of the gauge invariance of the expression 
in Eq.~(\ref{temp1}) has been 
already verified in Ref.~\onlinecite{gamma}.
The final expression reads as
\begin{eqnarray}
\tilde{E}_{x}^{\rm 1BSDM}  
& = & - \frac{\pi}{2} \sum_{\sigma=\uparrow,\downarrow} 
\int d^{2}r \left[ {\sqrt\pi} + \frac{3}{4} \tilde{A}_{\sigma} \right]
\nonumber \\ 
& \times & \rho^2_{\sigma}(\vr)\tilde{\beta}^{1/2}_{\sigma}(\vr)\,,
\label{AEX_4}
\end{eqnarray}
where $\tilde{A}_\sigma$ is determined through the normalization as
\begin{equation}
N_{\sigma} = \pi \int d^{2}r \left[ 1 + 2 \tilde{A}_{\sigma} \right] \rho^2_{\sigma}(\vr) \tilde{\beta}_{\sigma}(\vr)\,.
\label{Anorm4}
\end{equation}
Equation~(\ref{AEX_4}) together with Eq.~(\ref{Anorm4}) provide another
density functional for the exchange energy. 
Finally, the x-hole potential has a form
\be
\tilde{U}^\sigma_{x,{\rm 1BSDM}}(\vr)
 =  -\pi
 \left[ \sqrt{\pi} + \frac{3}{4} \tilde{A}_{\sigma} \right] \rho_{\sigma}(\vr)\tilde{\beta}^{1/2}_{\sigma}(\vr)\,.
\label{xholepot2}
\ee
Similarly to the previous section, the exchange-hole 
potentials and exchange energies
calculated {\em without} the modification
of $\tau_\sigma$ are denoted below simply without the tilde
symbols, i.e., ${U}^\sigma_{x,{\rm 1BSDM}}$ and
${E}^\sigma_{x,{\rm 1BSDM}}$.

\subsection{Correlation energy from the 
correlation-hole functions}\label{correlation}

High predictive power in the application of DFT
requires the accurate treatment of the electronic
correlation in both inhomogeneous systems and
in the limit of the homogeneous electron gas.
We have achieved this goal in 2D
by generalizing our previous approximation~\cite{c1} 
to a parameter-free form,~\cite{c2} which reproduces the
correlation energy of the 2DEG while preserving the
ability to deal with inhomogeneous systems (quantum dots).

The correlation energy is expressed in terms of the
cylindrical average of the (coupling-constant dependent) 
correlation-hole (c-hole) functions
$ \bar{h}^{\sigma\sigma'}_{c,\lambda}(\vr,s)$ as follows:~\cite{dft1,dft3}
\be
E^{\sigma\sigma'}_{c}= \pi  \int d\vr\, \rho_{\sigma}(\vr)  
\int_{0}^{\infty} ds \, 
\int_{0}^{1} d\lambda \, \bar{h}^{\sigma\sigma'}_{c,\lambda}(\vr,s)\;.
\ee
It is obvious that an approximation for 
$ \bar{h}^{\sigma\sigma}_{c,\lambda}(\vr,s)$ implies an
approximation for 
$E^{\sigma\sigma'}_{c}$. 
In modeling these quantities, we have proposed a form~\cite{c1} 
\be\label{mch1}
\bar{h}^{\sigma\sigma}_{c,\lambda}(\vr,s)  \approx
\frac{2\lambda s^2}{3}  
\left[ \frac{\left(s-z_{\sigma\sigma}(\vr) \right) D_{\sigma}(\vr) }{1+ \frac{2}{3}  \lambda z_{\sigma\sigma}(\vr) } \right]\exp{\left(-\frac{9\pi s^2}{16 z^2_{\sigma\sigma}(\vr) }\right)} 
\ee
\be\label{mch2}
\bar{h}^{\sigma{\bar \sigma}}_{c,\lambda}(\vr,s)  \approx
2 \lambda \rho_{{\bar \sigma}}(\vr)
\left[ \frac{s-z_{\sigma{\bar \sigma}}(\vr)}{1+ 2  \lambda z_{\sigma{\bar \sigma}}(\vr) } \right]
\exp{\left(-\frac{\pi s^2}{4 z^2_{\sigma\bar\sigma}(\vr) }\right)}  \;
\ee
for the same- and opposite-spin cases, $\sigma\sigma'=\sigma\sigma$
and $\sigma\sigma'=\sigma{\bar\sigma}$, respectively. 
Here
\be\label{D}
D_{\sigma}(\vr):= \frac{1}{2} \left( \tau_\sigma - \frac{1}{4} \frac{\left( \nabla \rho_\sigma 
\right)^2}{\rho_\sigma} - \frac{\vj^2_{p,\sigma}}{\rho_\sigma} \right)\;
\ee
and
\bea \label{z1}
z_{\sigma\sigma}(\vr) & := & 2 c_{\sigma\sigma}
|U_x^{\sigma}(\vr)|^{-1}, \\
z_{\sigma{\bar \sigma}}(\vr) & := & c_{\sigma{\bar \sigma}} \left[ |U_x^{\sigma}(\vr)|^{-1}+ |U_x^{{\bar \sigma}}(\vr)|^{-1}\right]\;.
\label{z2}
\eea
Equations~(\ref{z1}) and (\ref{z2}) are proportionality relations  
that may be enforced {\em locally} 
in space [see  Eqs.~(\ref{c_par1}) and (\ref{c_par2}) below].
It is apparent that $z_{\sigma\sigma'}(\vr)$ 
set the characteristic sizes of the c-hole functions in terms
of the sizes of the x-hole functions.
The idea behind this assumption is the following:
the smaller the x-hole around each electron is, the more tightly
the electrons are screened. Therefore, they are expected to be
correlated much less. 

The above modeling provides~\cite{c1}:
(i) zero correlation energy for one-particle systems (as the exact one);
(ii) exact short-range behavior of the $\lambda$-dependent c-hole 
functions; (iii) a ``reasonable'' decay in the limit $s \rightarrow \infty$; 
(iv) exact normalization
of the $\lambda$-dependent c-hole functions. Furthermore, 
(v) $c_{\sigma{\sigma'}}$ can be defined in such a way that the 
total correlation energy of the 2DEG 
is exactly reproduced.~\cite{c2} As a result,
when the (average) density has a realistic range, 
$0<r_s=1/\sqrt{\pi \rho}<20$,
we can use the following (approximate) 
parameterizations:
\be
\label{c_par1}
c_{\sigma\sigma}[r_s] = \alpha\,\log(r_s)+\beta\,r_s^{\gamma}
\ee
with $\alpha=-0.1415\,1$, $\beta=1.226\,1$, $\gamma=0.144\,99$,
and
\be\label{c_par2}
c_{\sigma{\bar \sigma}}[r_s] = \delta\,r_s^{\xi}
\ee
with $\delta=0.663\,25$ and $\xi=0.123\,96$.
When using the present correlation functional, the
coefficients $c_{\sigma\sigma'}[r_s](\vr)$  must be calculated 
at each point in space by making use of the density, that is, 
$r_s(\vr)= 1/\sqrt{\pi \rho(\vr)}$.

Here we propose a few simplifications along the 
lines of the previous sections. 
First, we may replace  $U_x^\sigma$ with an approximate 
expression obtained in Sec.~\ref{x-hole}. 
Secondly, we 
apply Eq.~(\ref{te})  to Eq.~(\ref{D}) leading to
\be\label{Dmod}
D_{\sigma}(\vr) \rightarrow \tilde{D}_{\sigma}(\vr)=  \pi \rho^2_\sigma(\vr) + \frac{1}{6} \nabla^2 \rho_\sigma(\vr) - \frac{1}{8} \frac{\left( \nabla \rho_\sigma(\vr) \right)^2}{\rho_\sigma(\vr)}\;.
\ee
Now, conditions (i) and (ii) given above are 
no longer satisfied. The latter modification clearly 
affects the {\em same-spin} c-hole functions. 
Applying both of the described simplifications -- which is naturally
required in order to make the functional orbital-free -- 
the correlation energy can be expressed in
terms of the (spin-dependent) c-hole potentials, 
$U^{\sigma\sigma'}_{c}(\vr)$, as follows:
\be
\label{Ec}
\tilde{E}^{\sigma\sigma'}_{c} = \frac{1}{2} \int d\vr\, \rho_{\sigma}(\vr)
\,\tilde{U}^{\sigma\sigma'}_{c}(\vr) \; ,
\ee
with
\bea \label{u1}
\tilde{U}^{\sigma\sigma}_c(\vr) & = & \frac{16}{81\pi}\left(8-3\pi\right) \tilde{D}_\sigma(\vr)
\tilde{z}^2_{\sigma\sigma}(\vr) \nonumber \\
& \times & \left[2 \tilde{z}_{\sigma\sigma}(\vr)-3\ln \left( \frac{2}{3}\tilde{z}_{\sigma\sigma}(\vr)+1\right)\right],
\eea
and 
\bea \label{u2}
\tilde{U}^{\sigma{\bar \sigma}}_c(\vr) & = & (2-\pi)\rho_{{\bar \sigma}}(\vr) \nonumber \\
& \times & \left[2
  \tilde{z}_{\sigma{\bar \sigma}}(\vr)-\ln\left(2 \tilde{z}_{\sigma{\bar \sigma}}(\vr) +1\right)  \right]\;,
\eea
where $\tilde{z}_{\sigma\sigma'}(\vr)$  are obtained 
by replacing $U^\sigma_x$ with $\tilde{U}^\sigma_x$ in
Eqs.~(\ref{z1}) and (\ref{z2}).

\section{Testing the modifications}\label{testing}

Next we test the modifications for different
2D quantum-dot systems.
As a standard test set we
consider parabolic (harmonic) dots consisting
of $N$ electrons confined in an 
external potential $v_{\rm ext}(r)=\omega^2 r^2 /2$.
First we use the {\tt octopus} code~\cite{octopus} 
to solve the KS problem self-consistently using
the exact-exchange (EXX) functional within the 
Krieger-Li-Iafrate (KLI) approximation.~\cite{KLI}
Then the resulting KS orbitals --   and for the modified
functionals solely the electron density and its gradients -- 
are used to compute the energy expressions and their 
ingredients introduced in the previous sections.
The obtained exchange energies can be directly compared
with the EXX-KLI results $E^{\rm EXX}_x$. In addition to this test set, 
we also consider a large quantum dot where we compare with
the LSDA, as well as a rectangular quantum slab.
In the case of correlation,
we exploit the numerically exact configuration-interaction
data~\cite{rontani} for the total energies $E^{\rm ref}_{\rm tot}$,
 so that the reference correlation energy can be evaluated from
$E_c^{\rm ref}=E^{\rm ref}_{\rm tot}-E^{\rm EXX}_{\rm tot}$.

\subsection{Exchange energies}

First we test the ingredients entering in
the expressions for the exchange energy in 
Secs.~\ref{x-hole} and \ref{1BSDMsec}. 

\begin{figure}
\includegraphics[width=0.8\columnwidth]{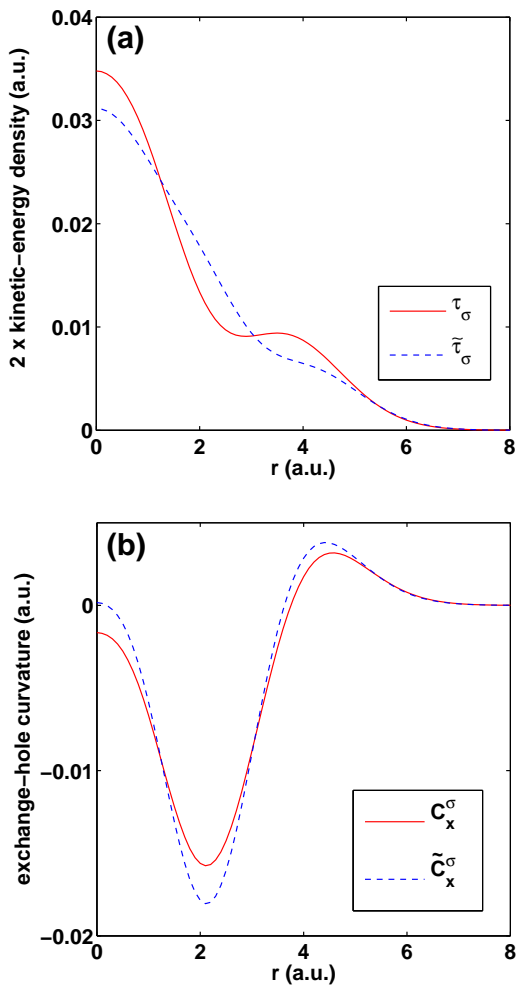}
\caption{(color online) (a) Comparison
of the original (solid line) and modified (dashed line) 
kinetic-energy density 
of a spin-polarized three-electron parabolic  ($\omega = 1/4$)
quantum dot. (b) Local curvature of
the exchange hole calculated from the original (solid line)
and modified (dashed line) kinetic-energy density.
}
\label{fig1}
\end{figure}

Figure~\ref{fig1}(a)
shows the original and modified kinetic-energy densities [as defined in Eq.~(\ref{te})]
of a spin-polarized three-electron parabolic quantum dot
with $\omega=1/4$. The characteristic step in $\tau_\sigma$
at the shell of the quantum dot at $r\sim 3$ is significantly 
smoother in ${\tilde \tau}_\sigma$.
However, this difference is partly washed away in the 
local curvature of the x-hole shown in Fig.~\ref{fig1}(b).

\begin{figure}
\includegraphics[width=0.8\columnwidth]{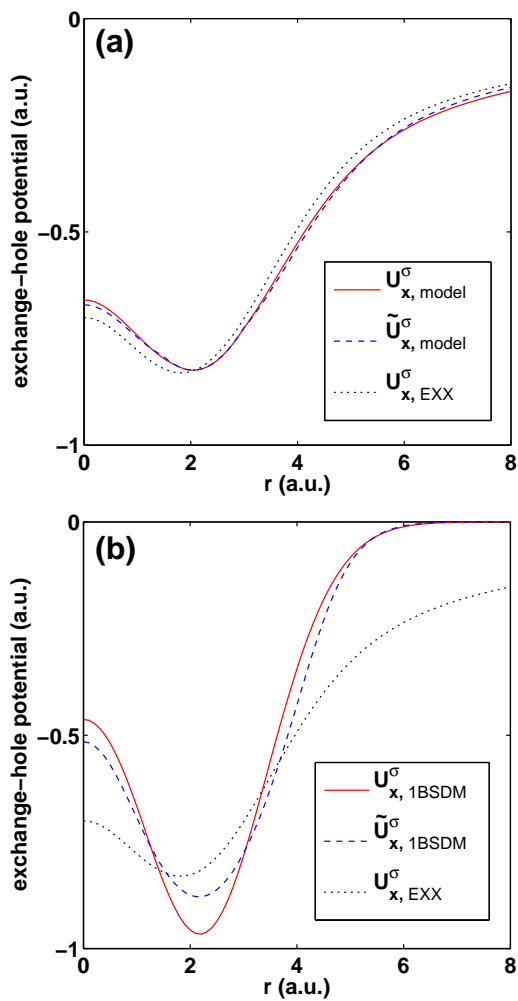}
\caption{(color online) 
Exchange-hole potentials of a spin-polarized 
three-electron parabolic ($\omega=1/4$)quantum dot. (a) Result
of the functionals in Ref.~\onlinecite{x1}
without (solid line) and with (dashed line) the 
modification (see Sec.~\ref{x-hole}).
(b) The same as (a) for the
functional in Ref.~\onlinecite{gamma} (see Sec.~\ref{1BSDMsec}).
The dotted line corresponds to the EXX-KLI
result.}
\label{fig2}
\end{figure}

In the exchange-hole potential shown in  Fig.~\ref{fig2}(a) 
-- computed with the functional described in Sec.~\ref{1BSDMsec}  --
the difference is reduced further, so that the
results are almost identical. Moreover, they
agree very well with the EXX-KLI result corresponding 
to the Slater potential (dotted line). Naturally, this
similarity leads to precise exchange energies as
explicitly shown below.

In Fig.~\ref{fig2}(b) we show the exchange-hole potentials
calculated with the functional
described in Sec.~\ref{1BSDMsec}.
Again, the results with and without the modification
of the functional are similar, although the relative
difference is larger than in the previous case [Fig.~\ref{fig2}(a)].
Both potentials, however, deviate rather strongly from the EXX-KLI
result. This tendency is already present in the original functional, which
has been tailored mainly to produce accurate exchange
{\em energies},~\cite{gamma} which is indeed the case as shown below.

Table~\ref{table1}
\begin{table}
\caption{\label{table1}
Exchange energies for fully spin-polarized parabolic 
quantum dots calculated using the functional 
of Ref.~\onlinecite{x1} ($E_x^{\rm model}$), its modification
described in Sec.~\ref{xhole} (${\tilde E}_x^{\rm model}$), the functional in
of Ref.~\onlinecite{gamma} ($E_x^{\rm 1BSDM}$), 
its modification (${\tilde E}_x^{\rm 1BSDM}$) described in Sec.~\ref{1BSDMsec},
and the local spin-density approximation ($E_x^{\rm LSDA}$.
They are compared with the EXX-KLI result ($E_x^{\rm EXX}$),
so that the last row shows the mean absolute error in percentage.
}
\begin{tabular*}{\columnwidth}{@{\extracolsep{\fill}} c c c c c c c c}
\hline
\hline
$N$& $B (T)$&  $E_x^{\rm model}$& ${\tilde E}_x^{\rm model}$& $E_x^{\rm 1BSDM}$& ${\tilde E}_x^{\rm 1BSDM}$& $E_x^{\rm LSDA}$& $E_x^{\rm EXX}$\\
\hline
2& 0&  -0.626&  -0.634&  -0.618& -0.620& -0.583&  -0.626 \\
3& 0&  -1.038&  -1.043&  -1.029& -1.021& -0.963&  -1.021 \\
3& 2&  -1.038&  -1.056&  -1.029& -1.037& -0.979&  -1.039 \\
4& 0&  -1.421&  -1.435&  -1.416& -1.408& -1.332&  -1.374 \\
5& 0&  -1.865&  -1.876&  -1.846& -1.842& -1.745&  -1.816 \\
6& 0&  -2.267&  -2.275&  -2.249& -2.241& -2.126&  -2.214 \\
6& 3&  -2.349&  -2.391&  -2.344& -2.362& -2.241&  -2.357 \\
\hline
 &  &   1.5  &   2.4  &   1.4  &  0.9   &  4.9 &  \\ 
\hline
\end{tabular*}
\end{table}
shows the exchange energies calculated for several quantum dots.
The set includes four cases with orbital currents (rows 1, 3, 4, and 7), 
in two cases arising from an external magnetic field perpendicular to the
2D plane (rows 3 and 7).~\cite{units}
We consider modifications for both $E_x^{\rm model}$ 
and $E_x^{\rm 1BSDM}$, respectively. Overall, the modifications
preserve the excellent performance of the functionals very well
(see the last row of Table~\ref{table1}). For the 1BSDM
approximation the modification even improves the performance.
The LSDA is giving clearly the worst accuracy of the tested
functionals.

\begin{figure}
\includegraphics[width=0.8\columnwidth]{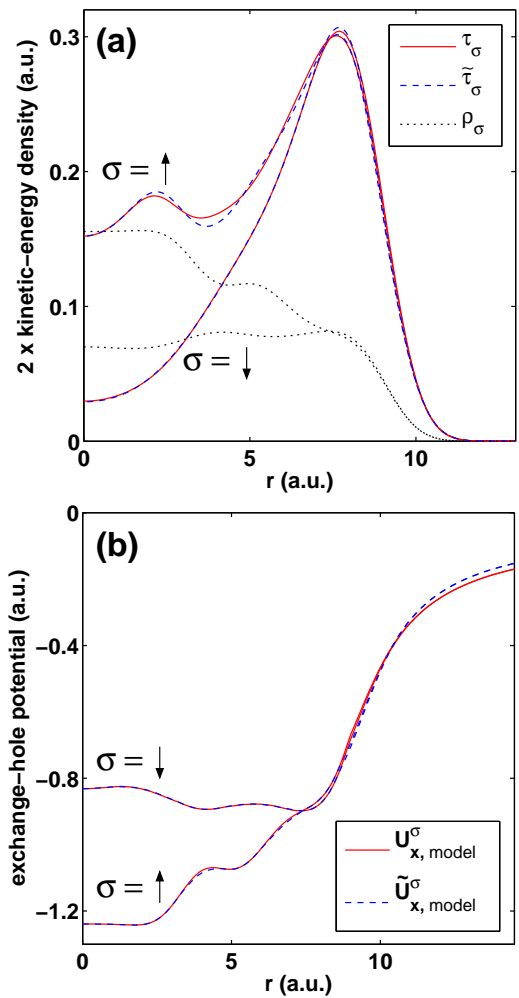}
\caption{(color online) 
(a) Original (solid lines) and modified (dashed lines) 
kinetic-energy densities for spin-up and spin-down 
electrons in a 48-electron parabolic ($\omega=0.3373$) quantum dot
at $B=3.05$ T. The dotted lines show the spin densities. 
(b) Resulting spin-up and spin-down 
exchange-hole potentials using
the functionals obtained from the modeling of the 
exchange hole (see Sec.~\ref{x-hole}). 
}
\label{fig3}
\end{figure}

In addition to the test set of Table~\ref{table1} that covers
only few-electron quantum dots, we now consider
two rather different cases.
First we focus on a large 48-electron parabolic  quantum dot
with $\omega=0.3373$ at a magnetic field of $B=3.05$ T.
This partially spin-polarized (total spin $S=3$) ground state 
has a compact ``spin droplet'' on the second-lowest Landau level,
and its existence has been confirmed in recent spin-blockade 
experiments.~\cite{rogge,droplet}
Here we have performed a LSDA calculation and use that density
as an input in the functionals.
Figure~\ref{fig3}
shows the kinetic-energy densities and exchange-hole
potentials calculated with the functionals 
obtained from the modeling of the 
exchange hole (see Sec.~\ref{x-hole}).
The original and modified $\tau_\sigma$ for both
spin-up and spin-down electrons are
very similar, and the resulting exchange-hole
potentials are practically the same.
The exchange energies are $E_x^{\rm model}=-22.18$ and 
${\tilde E}_x^{\rm model}=-22.25$ (difference of $0.3\%$).
In comparison, the LSDA yields $E_x^{\rm LSDA}=-21.11$.
In lack of a reliable EXX reference result for a system 
of this size it is not possible to
judge whether the exchange energy from the model(s)
or from the LSDA is more accurate.
However, knowing that the LSDA typically underestimates the
(absolute value of) $E_x$, our results for
$E_x^{\rm model}$ and ${\tilde E}_x^{\rm model}$ deviate
from the LSDA in the correct direction. Most importantly,
the exchange-hole potentials from the model are more 
accurate, especially in the asymptotic region.~\cite{x1}

\begin{figure}
\includegraphics[width=0.7\columnwidth]{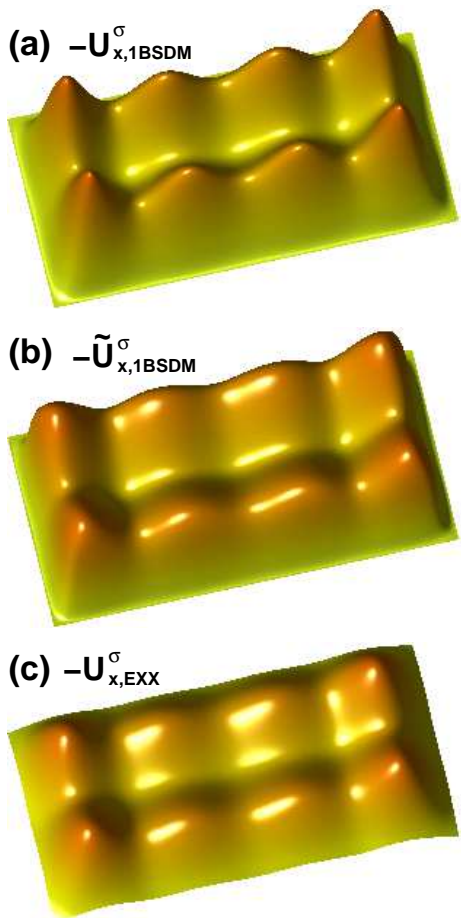}
\caption{(color online) 
Exchange-hole potentials of a 16-electron 
rectangular quantum dot calculated with
the original (a) and modified (b) functionals 
obtained from the one-body spin-density matrix
(see Sec.~\ref{1BSDMsec}) in comparision with
the exact-exchange (Slater) potential (c).
}
\label{fig4}
\end{figure}

Our second example of a quantum-dot system that differs
from those in Table~\ref{table1} is a 16-electron rectangular
hard-wall quantum slab with size $2\sqrt{2}\pi \times \sqrt{2}\pi$
corresponding to $\sim 90\,{\rm nm}\times 45\,{\rm nm}$ 
in SI units.~\cite{units} In Fig.~\ref{fig4}
we compare the exchange-hole potentials given by
the 1BSDM functionals (see Sec.~\ref{1BSDMsec}) 
to the EXX-KLI (Slater) potential computed. The overall 
shapes are very similar, but as expected, the EXX-KLI potential
is considerably smoother. Interestingly, however, the modified
potential is qualitatively closer to the EXX-KLI result than the 
original one. Regarding the exchange energies both functionals
perform similarly: $E_x^{\rm 1BSDM}=13.13$,
${\tilde E}_x^{\rm 1BSDM}=13.15$, and  $E_x^{\rm EXX}=12.7$.
Thus, the modification in $\tau_\sigma$ is well justified also when
considering a 2D system with a hard-wall geometry.

\subsection{Correlation energies}

\begin{figure}
\includegraphics[width=0.8\columnwidth]{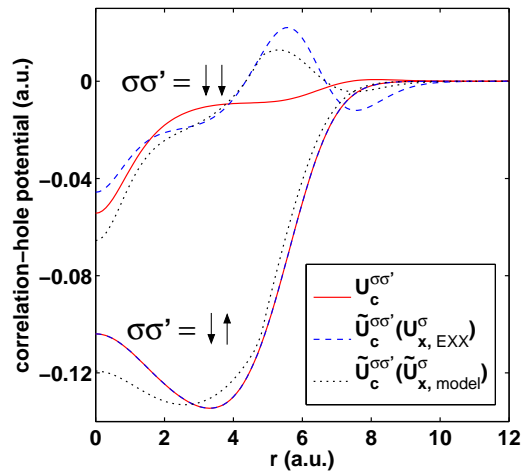}
\caption{(color online) 
Spin-pair components of the correlation-hole potential 
for a spin-unpolarized 
six-electron parabolic ($\omega = 1/4$) quantum dot calculated using the 
functional obtained from the correlation-hole 
modeling~\cite{c2} (solid  line), 
its modification described in Sec.~\ref{correlation} 
with $U^\sigma_{x,{\rm EXX}}$ (obtained within the KLI approximation) as an input (dashed line),
and with ${\tilde U}^\sigma_{x,{\rm model}}$ as an input (dotted line).
}
\label{fig5}
\end{figure}

Figure~\ref{fig5}
shows the spin-pair components of the correlation-hole potentials
for a spin-unpolarized (total spin $S=0$) six-electron parabolic 
quantum dot ($\omega=1/4$) calculated with the model 
of Ref.~\onlinecite{c2} in comparison with its modifications
introduced in Sec.~\ref{correlation}. The modification
induces clear devitations from the original potential,
especially for the same-spin component affected by
modifications in $D_\sigma$. For example, the bump at $r\sim 4$ is
due to the change of sign in ${\tilde D}_\sigma$ in
that regime. In contrast, the opposite-spin 
component is independent of $D_\sigma$ [see Eq.~(\ref{u2})],
so that the modified functionals are almost the same;
here the choice of ${\tilde U}^\sigma_{x,{\rm model}}$ instead
of $U^\sigma_{x,{\rm EXX}}$ has a negligible effect (solid
and dashed lines overlap). In lack of an
{\em exact} reference result we cannot assess the quality 
of the correlation-hole potential(s). Hence, in the following we will
focus on the correlation energies for which reference
results can be obtained as described in the beginning
of Sec.~\ref{testing}.

In Table~\ref{table2}
\begin{table}
\caption{\label{table2}
Correlation energies for spin-polarized ($S=N/2$) 
and unpolarized ($S=0$)
parabolic quantum dots calculated using the functional 
in Ref.~\onlinecite{c2} (see Sec.~\ref{correlation}), 
its modification with $U^\sigma_{x,{\rm EXX}}$ (in the KLI approximation) as an input,
the modified form with ${\tilde U}^\sigma_x$ as an input, and 
the local-spin-density approximation.~\cite{LSDA_explain}
The last column shows the numerically exact reference result.
The last row shows the mean absolute error in percentage.
}
\begin{tabular*}{\columnwidth}{@{\extracolsep{\fill}} c c c c c c c c}
\hline
\hline
$N$& $S$ & $\omega$&  $E_c$& ${\tilde E}_c(U^\sigma_{x,{\rm EXX}})$& ${\tilde E}_c({\tilde U}^\sigma_{x,{\rm model}})$& $E_c^{\rm LSDA}$& $E_c^{\rm ref}$\\
\hline
2 &  1   &1/4  &     -0.0115 &  -0.0073&   -0.0085&   -0.0345 &  -0.0100 \\
3&   3/2 &1/4  &     -0.0225 &  -0.0189 &  -0.0200&   -0.0564&   -0.0226 \\
4 &  2   &1/4  &     -0.0399 &  -0.0337 &  -0.0330&   -0.0730 &  -0.0337 \\
5 &  5/2 &1/4   &    -0.0570 &  -0.0465 &  -0.0468&   -0.0929&   -0.0484 \\
6 &  3   &1/4   &    -0.0681 &  -0.0617&   -0.0613&   -0.1125 &  -0.0640 \\
6 &  0   &1/4   &    -0.390  &  -0.379 &   ­0.380 &   -0.458  &  -0.396  \\
3 &  3/2 &1/16  &    -0.0138 &  -0.0106&   -0.0122&   -0.0382 &  -0.0167 \\
5 &  5/2 &1/16  &    -0.0348 &  -0.0278&   -0.0299&   -0.0659 &  -0.0357 \\
6 &  3   &1/16   &   -0.0426  & -0.0372&   -0.0393&   -0.0796 &  -0.0459 \\
6 &  0   &1/16   &   -0.228  &  -0.214 &   -0.220&    -0.282  &  -0.250  \\
\hline
 &  &   &          9.5  &   15  &   11  &  99  &  \\ 
\hline
\end{tabular*}
\end{table}
we test the effect of the modifications for the
accuracy of the correlation-energy functional. As discussed at the end of 
Sec.~\ref{correlation}, we consider two approximations, where we
either do {\em not} approximate $U_x$ in the same framework
but use the EXX result, or then we apply the modification
also to $U_x$. Interestingly, the best result -- apart from 
the original functional which is very close in accuracy --
is given by ${\tilde E}_c({\tilde U}^\sigma_x)$ for both
spin-polarized and unpolarized cases. This finding
may demonstrate the compatibility between the corresponding
exchange- and correlation-energy functionals. Nevertheless,
all the functionals introduced here are superior to the LSDA,
whose error is an order of magnitude larger.~\cite{c1,c2}

\begin{figure}
\includegraphics[width=0.8\columnwidth]{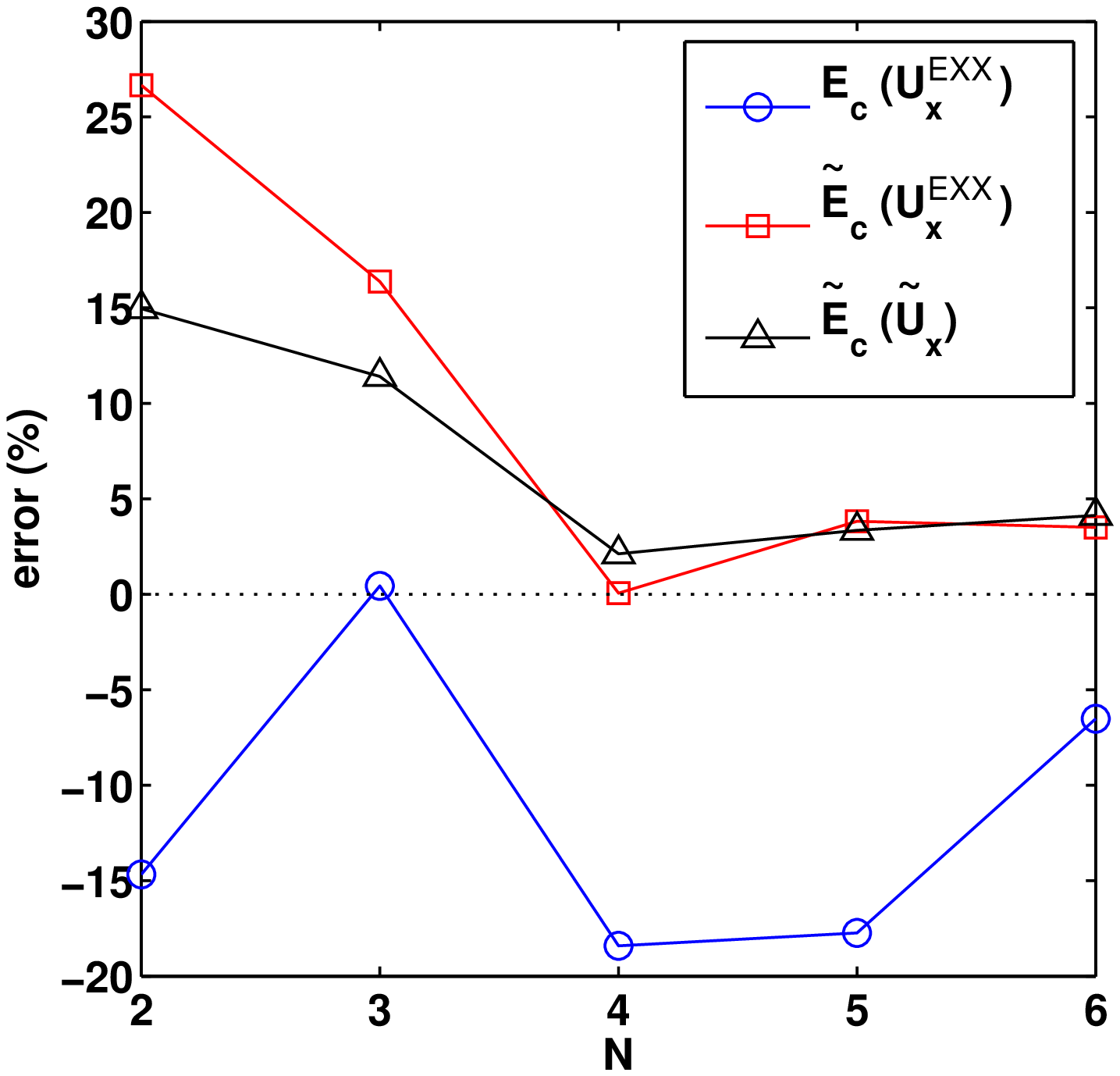}
\caption{(color online) Error in the correlation
energy of spin-polarized parabolic  ($\omega=1/4$)  quantum dots
with $N$ electrons
obtained using different approximations.
The circles correspond to the functional
of Ref.~\onlinecite{c2} and
the squares show the performance of the modified 
functional introduced in Sec.~\ref{correlation}, when
the exact exchange-energy potential has been used 
in the expression. The triangles 
correspond to the case when also the 
exchange-energy potential has been used in
a (similar) modified form.
}
\label{fig6}
\end{figure}

Finally, in Fig.~\ref{fig6}
we plot the relative errors of the correlation-energy
functionals as a function of $N$ (cf.
Fig. 2 in Ref.~\onlinecite{c2}). It is interesting
to note that, at least for this set of systems,
the modified functionals show consistent behavior as
a function of the number of electrons. Thus, it may
be expected that the good performance continues further
to larger $N$. Unfortunately, a throughout testing of
this is beyond the capability of numerically exact
methods to provide accurate reference data.

\section{Conclusions and outlook}

In this work we have explored the possibility to
modify meta-generalized-gradient approximations (meta-GGAs)
for the exchange and correlation energies of two-dimensional
systems to Laplacian-level meta-GGA ones.
We have analyzed the effects of the according 
modifications on various systems. 
Although the differences in the kinetic-energy densities 
can be considerable,
the functionals considered in this work
preserve well the quality of the exchange- and 
correlation-hole potentials, and in particular the
corresponding energies. Overall, we find that the
performance is well preserved, if not even improved, 
by the modifications. 

The simplified meta-GGAs provide significant practicality
and numerical efficiency in the application of the functionals.
Therefore, a self-consistent and multi-purpose implementation
of the present toolbox of functionals is now within reach, 
enabling the investigation of (quasi-)two dimensional 
electronic nanostructures of experimental and technological relevance.

\begin{acknowledgments}
This work has been  supported by 
DOE grant DE-FG02-05ER46203 (S.P.) 
and by the Academy of Finland (E.R.).
\end{acknowledgments}


\begin{thebibliography}{ll}
\bibitem{units} When transforming the (effective) atomic units given in the paper
to SI units, we use the effective-mass approximation with the 
typical GaAs parameters: $m^\ast=0.067\,m_0$ and $\varepsilon=12.4\,\varepsilon_0$.
Hence, the energies, lengths, and magnetic-field strengths scale as 
$E_{h}^\ast=(m^\ast/m_0)/(\varepsilon/\varepsilon_0)^2E_h\approx 12\,\mathrm{meV}$,
$a_0^\ast=(\varepsilon/\varepsilon_0)/(m^\ast/m_0) a_0\approx 10\,\mathrm{nm}$, and
$B_0^\ast=(m^\ast/m_0)^2/(\varepsilon/\varepsilon_0)^2 B_0 \approx 6.9\,\mathrm{T}$, respectively.
In the paper the magnetic fields are always given in Tesla for clarity.
\bibitem{LSDA_explain} Note that here the LSDA correlation 
energies have been calculated from the self-consistent EXX-KLI densities, 
whereas in Refs.~\onlinecite{c1} and \onlinecite{c2} 
the LSDA correlation energies have been obtained self-consistently. 
The differences between the results are small.
\end{thebibliography}
\end{document}